\documentclass[]{aa}
\usepackage{graphicx}
\usepackage{txfonts}
\usepackage{natbib}
\usepackage{pifont}
\newcommand{\cmark}{\ding{51}}
\newcommand{\xmark}{\ding{55}}
\usepackage{hyperref} 
\usepackage{dsfont}
\usepackage{amssymb}
\usepackage[dvipsnames]{xcolor}
\hypersetup{colorlinks=true,linkcolor=blue,citecolor=blue,urlcolor=blue}
\usepackage{orcidlink}
\usepackage{lscape}
\begin{document}

%\usepackage{txfonts}

%\setcitestyle{citesep={,}}

%%%%% AUTHORS - PLACE YOUR OWN PACKAGES HERE %%%%%

% Only include extra packages if you really need them. Common packages are:
        % Including figure files
% Extra maths symbols

%%%%%%%%%%%%%%%%%%%%%%%%%%%%%%%%%%%%%%%%%%%%%%%%%%

\title{Tracing the Milky Way spiral arms with $^{26}$Al.} \subtitle{The role of nova systems in the 2D distribution of $^{26}$Al}

%%%%%%%%%%%
\author {A. Vasini \orcidlink{0009-0007-0961-0429} \inst{1,2} \thanks{email to: arianna.vasini@inaf.it}, E. Spitoni \orcidlink{0000-0001-9715-5727}\inst{2},  F. Matteucci \orcidlink{0000-0001-7067-2302} \inst{1,2,3}    G. Cescutti \orcidlink{0000-0002-3184-9918}   \inst{1,2,3},\and M. Della Valle \orcidlink{0000-0003-3142-5020} \inst{4,5}  }
\institute{Dipartimento di Fisica, Sezione di Astronomia,
  Universit\`a di Trieste, Via G.~B. Tiepolo 11, I-34143 Trieste,
  Italy \and I.N.A.F. Osservatorio Astronomico di Trieste, via G.B. Tiepolo
 11, 34131, Trieste, Italy  
   \and I.N.F.N. Sezione di Trieste, via Valerio 2, 34134 Trieste, Italy
   \and INAF - Osservatorio Astronomico di Capodimonte, Salita Moiariello 16, I-80131, Napoli, Italy 
\and ICRANet, Piazza della Repubblica 10, I-65122 Pescara, Italy
 }

 \date{Received xxxx / Accepted xxxx}

\abstract {Massive stars are one of the most important and investigated astrophysical production sites of $^{26}$Al, a short-lived radioisotope with $\sim$ 1 Myr half-life. Its short lifetime prevents us from observing its complete chemical history, and only the $^{26}$Al that was recently produced by massive stars can be observed. Hence, it is considered a tracer of star formation rate (SFR). However, important contributions to $^{26}$Al comes from nova systems that pollute the interstellar medium  with a large delay, thus partly erasing the correlation between $^{26}$Al and SFR.}
{In this work we describe the 2D distribution of the mass of $^{26}$Al as well as that of massive stars and nova systems in the Milky Way, to investigate their relative contributions to the production of $^{26}$Al.}
%{In this work, we aim at describing the 2D distribution of the mass of $^{26}$Al as well as nova systems and massive stars in the Milky Way, and quantifying the relative contributions of massive stars and nova systems to the production of $^{26}$Al.}
{We use a detailed 2D chemical evolution model where the SFR is azimuthally dependent and is required to reproduce the spiral arm pattern observed in the Milky Way. We test two different models, one where the $^{26}$Al comes from massive stars and novae, and one with massive stars only. We then compare the predictions to the $\sim$ 2 M$_{\odot}$ of $^{26}$Al mass observed by the surveys COMPTEL and INTEGRAL.}
{The results show that novae do not trace SFR and, in the solar vicinity, they concentrate in its  minima. The effect of novae on the map of the $^{26}$Al mass consists in damping the spiral pattern by a factor of five. Regarding the nucleosynthesis, we find that $\sim$75\% of the $^{26}$Al is produced by novae and the $\sim$25\% by massive stars.}
{We conclude that novae cannot be neglected as $^{26}$Al producers since the observations can only be reproduced by including their contribution. Moreover, we suggest that bulge novae should eject around six times more material than the disc ones to well reproduce the observed mass of $^{26}$Al.}
%{We conclude that novae cannot be neglected as $^{26}$Al producers {\bf qui bisognerebbe dare una percentuale per il contributo delle novae} and that the bulge novae should eject ten times more material than the disc ones. Under these assumptions, we obtain a total mass of 2.882 M$_{\odot}$ of $^{26}$Al within 5 kpc from the Galactic center, in agreement with observations.}
%{In conclusion, we stress that, because of  the production of  $^{26}$Al by novae, this element is not a good tracer of the star formation rate in the Milky Way disc; in particular, the contribution from novae lower by a factor of 4 the correlation with the star formation rate and that to recover the observed $^{26}$Al mass, bulge novae are required to pollute a factor of ten more than disc novae.}
%{\bf COMMENTO: NON SI METTONO ACRONIMI NELL' ABSTRACT. L'ABSTRACT ERA RIPETITIVO E L?HO RIFORMULATO}


\titlerunning{Effects of multiple spiral arm patterns  on abundance gradients}

\authorrunning{Vasini et al.}

\maketitle

\section{Introduction}
\label{sec:introduction}

Short-lived radioisotopes (SLRs) are a class of nuclei that undergo a radioactive decay occurring on very short astronomical timescales ($\sim$1-10 Myr). They represent the link between different approaches to unveil the processes behind Galactic evolution, from chemical evolution \citep{Timmes+95} to the measurements of the isotopical abundances in meteorites coming from the solar neighbourhood \citep{Desch+2023}. Other theoretical methods tested were the simulations of the superbubbles \citep{Schulreich+23}, Galactic population synthesis models \citep{Siegert+23} and the stochastic chemical evolution studies such as those presented in  \citet{cote+19_giugno,cote+19_dicembre}. Given their brief half-life, these nuclei rapidly decay into the final product of their decay chain, leaving behind no trace of their previous presence. Therefore, all the measurements and the theoretical studies addressed to SLRs focus on the present day scenario only, because no heritage of the past history is left.

The astronomical interest behind $^{26}$Al lies also in its close relation to the star formation rate (SFR). %The centrality of this issue is even clearer when considering the role of $^{26}$Al in chemical evolution and observations. 
Since, generally, massive stars have been considered as the main source of $^{26}$Al, this isotope is restored into the insterstellar medium (ISM) shortly after the birth of its stellar progenitor. Moreover, its short decay timescale prevents $^{26}$Al from moving too far away from the location of the progenitor. As a consequence, since massive stars are located in the active star formation regions of the Galaxy, this isotope will also be detected in the proximity of these areas. Hence, $^{26}$Al has been  considered a good tracer of star formation regions \citep{Limongi+06}. 

In the past, chemical evolution studies relative to SLRs, tried to reproduce the observations provided in the last decades by the $\gamma$-ray astronomy community (\citealt{Vasini+22,Wehmeyer+23}) and offered predictions for upcoming observations of external structures \citep{vasini2023}.
%In this context, mono-dimensional models have always been the strategy followed to carry out these analyses. 

Many of these results were obtained by means of 1D chemical evolution models which assume that every observable (such as the SFR, the infall rate, the gas and stellar masses and the abundance gradients) depends only on the radial coordinate. Hence, the Milky Way (MW) is modeled by assuming a concentric ring structure, where the gas in each ring instantaneously and homogeneously mixes (instantaneous mixing approximation). %Therefore the only inhomogeneity considered is the exponential decay of the surface gas mass density going from the innermost regions towards the outskirts. 
%In a mono-dimensional chemical evolution model this exponential is recovered by assuming an annular concentric structure and requiring that the gas within each ring instantaneously and homogeneously mixes. 
%This therefore imply that the gas within each ring instantaneously and homogeneously mixes. 
%This mixing is indeed occurring in the ISM on a timescale which is %considered negligible in the majority of chemical evolution models. RIPETIZIONE

The picture becomes more complex when dealing with SLRs since an additional timescale is introduced, namely the timescale of their radioactive decay, and in this particular case that of $^{26}$Al ($\sim$ 1 Myr, see \citealt{Diehl+13}). 
$^{26}$Al, after being re-injected into the ISM, starts spreading in the surrounding regions as stable isotopes also do but, simultaneously, it also decays into $^{26}$Mg. As a consequence, the $^{26}$Al atoms will not mix homogeneously, because the mixing timescale is very likely longer than the $^{26}$Al decay timescale.
%will never reach a configuration where they are distributed along the whole annulus because the time necessary to reach the homogeneity is much longer than the decay timescale. 
Therefore, for $^{26}$Al, such as for other SLRs, the assumption of homogeneous mixing does not hold anymore. 
This fact would not be a problem if the SFR in the Galaxy was homogeneous but, as observations confirm, the SFR is concentrated in the spiral arms of the MW. Therefore a 1D chemical evolution model that assumes homogeneous mixing is not suitable for dealing with SLRs in the Galaxy. The solution is introducing a second dimension, namely the azimuth, in order to trace inhomogeneities in each ring.

Moreover, the scenario is further complicated by the part played by the nova systems, that represent a second but non negligible source of $^{26}$Al \citep{Nofar+91}. The chemical enrichment from novae is characterized by a long  time delay due to the combination of the time necessary to form a white dwarf and the additional cooling time requested to trigger the nova explosion \citep{DAntona&Matteucci91}. The direct consequence of this is that novae, unlike massive stars, do not trace the SFR. Given that a fraction of the $^{26}$Al in the Galaxy has a nova origin, the $^{26}$Al-SFR correlation is necessarily reduced, even though it is not yet clear how much.
To determine the damping due to the nova contribution, a 2D chemical evolution model is needed. 

%\textbf{1D chemical evolution models are based on an assumption that for SLRs is too strong, hence a 2D model where this assumption is relaxed represents a better choice.}

%For a stable isotope this assumption does not affect the results much. The results obtained through the decades with 1D chemical evolution models show a notable level of agreement with the observations and the ultimate confirmation was provided when two-dimensional chemical evolution models were developed. 

\citet{Spitoni+19,Spitoni+23} proposed a 2D chemical evolution model where azimuth and radial coordinates were considered. The azimuth is introduced by perturbing the SFR so as to reproduce the MW spiral arms. In these papers, every ring is divided into smaller cells, and the homogeneity is not imposed over the whole ring but rather over the single cells. This updated scenario 
allows the authors to show clearly that the 1D models represent a valid alternative almost everywhere in the Galaxy, except when dealing with the co-rotational radii regions where oscillations in chemical abundances over cells in the same ring arise and only the 2D model can trace them.

From the observational side, evidences of SLRs in the Galaxy have been collected adopting $\gamma$-ray astronomy techniques. Since 1991 two surveys, at first COMPTEL (\citealt{Schonfelder+84,Diehl+95}) and later INTEGRAL (\citealt{Winkler+94,Pleintinger+23}, have been dedicated to exploring the distribution of these elements along different longitudes on the Galactic plane and different latitudes above and below the Galactic plane. Among the most important results that were provided there is the 2D cartography of the $^{26}$Al in the MW (see \citealt{Pluschke+01,Diehl&Prantzos23}) where the evidence of a diffuse emission on the Galactic plane is reported together with some localized star formation regions superimposed. From the same data, also integrated measurements of the mass of $^{26}$Al in the region included within 5 kpc from the Galactic center were provided. {\citet{Prantzos&Diehl96}, adopting the flux measurements from COMPTEL, estimated an $^{26}$Al mass between 1.5 and 2 M$_{\odot}$ within 5 kpc from the Galactic center, whereas by making use of the measurements from INTEGRAL \citet{Diehl+16} proposed 2.0 $\pm$ 0.3 M$_{\odot}$ in the same 5 kpc radius ring. In \citet{Vasini+22} we already provided a theoretical estimate of $^{26}$Al integrated over that same region by adopting a 1D chemical evolution model to reproduce the disc and the bulge separately. With the prescriptions adopted, we were able to reproduce the observations and constrain the stellar yields, showing that a non negligible contribution from novae is necessary. A yet unexplored question is how much the integrated values are influenced by the adoption of a 1D rather than a 2D model. A 2D model, as already explained above, would be able to trace the oscillations in the SFR across the MW, at variance with a 1D model. Moreover, regarding the integrated values within the region of interest no conclusions can be drawn until a suitable 2D chemical evolution model for radioactive isotopes is developed. 
\newline

With this work we want to provide a 2D chemical evolution model to follow the evolution in space (as a function of radial distance and azimuth) and time of $^{26}$Al in the MW by adding an azimuth-dependent perturbation to the SFR that resembles the spiral arm pattern observed. Our aim is to showcase the impact of the nova population on the present day distribution of this isotope. Moreover, with our model we can hopefully have deeper insights on the differences between the novae in the bulge and in the disc, by modeling these two regions separately.

The paper is organized as follows. In Sec. \ref{sec:1D_model} we present the chemical evolution model adopted starting from the original 1D version for thick and thin discs and the bulge, and then we move to the nucleosynthesis considered with the description of the two models tested. In Sec. \ref{sec:2D_model} we show how we extended the 1D model by adding the azimuthal dependency to the SFR, then in Sec. \ref{sec:constraints} we show the Galactic observational constraints that we can reproduce. We present our results about novae and $^{26}$Al in Sec. \ref{sec:result_all} and, finally, we summarize our main conclusions in Sec. \ref{sec:conclusions}.

\section{The reference 1D chemical evolution model}
\label{sec:1D_model}

In this Section, we present the reference 1D chemical evolution model that we use as starting point for this work.  The 2D extension will be described later in Sec. \ref{sec:2D_model}.

We use as reference model the one adopted in \citet{Vasini+22}. The Galaxy is composed of two separated regions, the bulge (up to 2 kpc from the Galactic center) and the disc (from 2 kpc to 15 kpc from the MW center), itself composed of two chemically distinct substructures, the thick disc and the thin disc.

The thin disc is divided in 1 kpc wide concentric rings, that are isolated from each other since we neglect every radial gas flow. The bulge as well is modeled by assuming a central region until 1 kpc from the center and a surrounding ring, from 1 kpc to 2 kpc. Moreover, in the 1D model each ring is chemically homogeneous.

\subsection{Galactic thick and thin discs}

We assume that thick and thin discs form out of gas coming from two different infall episodes according to:
\begin{equation}
    A(R,t)=a(R)e^{-t/\tau_{T}} + b(R)e^{-(t-t_{max})/\tau_{D}(R)},
    \label{eq:two_infall}
\end{equation}
where the first episode is responsible of the formation of the thick disc on a timescale $\tau_{T}$=1.0 Gyr, whereas the thin disc is formed by the second infall following an "inside-out" scenario (see \citealt{matteucci1989}). To reproduce that, the second infall timescale is dependent on the Galactocentric distance following the law $\tau_{D}(R) = 1.033\,\, (R/\text{kpc}) - \text{1.267} $ Gyr (\citet{Chiappini+2001}). 
Additionally, $t_{\text{max}}=1 \, \text{Gyr}$ represents the time of maximum infall in the second accretion episode, indicating the delay between the end of the first and the start of the second infall episode. 
The parameters $a(R)$ and $b(R)$  are adjusted to match the current total surface mass density of the thick and thin discs, set to 17 M$_{\odot}\,\text{pc}^{-2}$ and 54 M$_{\odot}\,\text{pc}^{-2}$, respectively \citep{Chiappini+2001}. 
%We assumed a total present-day surface mass density in the solar vicinity of $\sigma_{tot}(t_G) = 54 \, M_{\odot} \, \text{pc}^{-2}$ \citep{Romano+2000}. 
%The parameters $\tau_T$ and $\tau_D(R)$ represent the timescales of the two infall episodes, expressed in Gyr. The first timescale, $\tau_T$, is associated with the formation of the thick disc, while the second, $\tau_D(R)$, is related to the thin disc and varies with Galactocentric radius. Based on previous studies, we assumed $\tau_T \text{Gyr}$ and calculated $\tau_D(R)$ using the relation $\tau_D(R) = 1.033 \cdot (R/\text{kpc}) - 1.267 \, \text{Gyr}$ to account for the 'inside–out' formation scenario \citet{Chiappini+2001}. 
\color{black}
A similar mechanism for the growth of the Galactic assembling mass is identified in cosmological simulations \citep{brook2012,bird2013,vincenzo2020}.

The SFR is expressed as the \citet{Kennicutt98} law:
\begin{equation}
    \psi(R,t) \propto \nu(R) \Sigma_{gas}^{k}(R,t),
    \label{eq:SFR}
\end{equation}
where $\nu(R)$ is the star formation efficiency that we assume to be 1.0 Gyr$^{-1}$ everywhere in the Galaxy, the same for thick and thin discs and $\Sigma_{gas}$ is the surface gas mass density. 

We assume a \citet{kroupa1993} IMF with three slopes, defined in a mass interval up to 100 M$_{\odot}$: 
\begin{equation}
\varphi_{\rm K93}(M)\propto \left \{ \begin{array}{rl}
M^{-1.3}\,\,\,\,\,\,\,\,\,\,\,\,\,\,\,\,\,\,\,\,\,\,\,\,\,\,\,\,\,\,M<0.5 M_{\odot}\\
M^{-2.2}\,\,\,\,\,\,\,\,\,\,\,\,\,\,\,\,\,\,0.5<M/M_{\odot}<1\\
M^{-2.7}\,\,\,\,\,\,\,\,\,\,\,\,\,\,\,\,\,\,\,\,\,\,\,\,\,\,\,\,\,\,\,\,\,M>1 M_{\odot}.
\end{array}
\right.
\label{eq:kroupa93}
\end{equation}

Regarding the stellar contribution to the ISM chemical composition, our model accounts for stars in different mass ranges, hence with different explosion mechanisms and lifetimes. 

The stars from M$_L$=8 M$_\odot$ up to M$_U$=30 M$_\odot$ die as Type II Supernovae (Type II SNe) whereas the more massive ones, from M$_L$=30 M$_\odot$ up to M$_U$=100 M$_\odot$ die as Type Ib SNe or Type Ic SNe.  These SNe are all core-collapse ones and their rate can be written as:
\begin{equation}
R_{\rm SNII,Ib,Ic}(t)=\int^{M_{\rm U}}_{M_{\rm L}} \psi(t-\tau_{\rm m})\varphi(m)dm,
    \label{eq:CC-SN}
\end{equation}
where the $\tau_m$ is the lifetime of a star of mass m.

On the other hand, the Type Ia SNe follow the single degenerate scenario from \citet{matteucci1986} and \citet{Matteucci_Recchi2001}, with the following rate:
\begin{equation}
R_{\rm SNIa}(t)=A_{\rm B}\int^{M_{\rm BM}}_{M_{\rm Bm}} \varphi(m)\bigg[\int^{0.5}_{\mu_{\rm Bmin}} f(\mu_{\rm B})\psi(t-\tau_{\rm m2})\bigg]dm.
    \label{eq:SNIa}
\end{equation}

The parameters are set as follows. A$_B$ is the fraction of binary systems that can generate a Type Ia SNe progenitor and it depends on the chosen IMF. In our case A$_B$ = 0.09. The mass $m$ over which the integration is performed is the total mass of the binary system, so as M$_{\rm Bm}$ and M$_{\rm BM}$ are the smallest and the largest possible total mass of the system, respectively. To generate a SNIa the two companions should be massive enough to make the white dwarf reach the Chandrasekhar mass, therefore we set M$_{\rm Bm}$ = 3 M$_{\odot}$. M$_{\rm BM}$ is instead set to 16 M$_\odot$ because the upper limit to form a CO white dwarf is 8 M$_\odot$. $f(\mu_{B})$ is the distribution of the mass ratio between the two stars, and $\psi(t-\tau_{m2})$ is the SFR at the time of formation of the secondary star, which represents the clock of the system.

Moreover, we also consider the contribution from nova systems, that are of particular interest in this case. They are implemented according to \citet{DAntona&Matteucci91}:
\begin{equation}
R_{\rm novaout}(t)=10^{4}\cdot\alpha\int^{8}_{0.8}\psi(t-\tau_{\rm m}-\Delta t)\varphi(m)dm,
\label{eq:nove}
\end{equation}
where $\alpha$ is the fraction of binary systems that are able to generate a nova, that was set to 0.0115. With this value we are able to reproduce the current nova rate in the MW disc. $\psi(t-\tau_m-\Delta t)$ is the SFR at the formation of the binary system. The delay between the nova explosion and its birth is given by the sum of two contributions, $\tau_m$, its lifetime, and the time needed for the white dwarf to cool down, $\Delta t$, set to 1.0 Gyr. In addition, the factor $10^{4}$ accounts for the several explosions that each nova system undergo during its lifetime (see \citealt{ford78}). The explosions of a nova system are distributed in time and should be modeled accordingly. This aspect has already been analysed by \citet{Cescutti&Molaro19} that improved the scenario by adopting multiple explosions, but did not find significant differences.

\subsection{The bulge}

Assuming that the Galactic bulge resides in the innermost region, within 2 kpc from the Galactic center, we model it separately, following the prescriptions from \citet{Matteucci+2019}. We divide it into a disc, from the center up to 1 kpc, and an external ring, from 1 kpc to 2 kpc, so that the concentric rings throughout the entire Galaxy are all equally thick. 

The gas forming the bulge is all accreted from a single infall episode according to:
\begin{equation}
    A(R,t)=a(R)e^{-t/\tau_B}
\end{equation}
where the timescale $\tau_B$ is set to 0.1 Gyr \citep{Ballero+07}. $a(R)$, as in the case of the disc, is a free parameter whose value is chosen to reproduce the mass density of the inner and the outer bulge regions, $\sim$ 1970 M$_{\odot}\,\text{pc}^{-2}$ and $\sim$ 1350 $M_{\odot}\,\text{pc}^{-2}$ , in agreement with \citet{Ballero+07} and \citet{Grieco+12}. 

Concerning the SFR, we adopt the same law as  in \ref{eq:SFR} with an efficiency $\nu$ of 25.0 Gyr$^{-1}$ and 35.0 Gyr$^{-1}$ for the ring and the inner region, respectively. This large star formation efficiency, together with the short timescale of the SFR, reproduces the rapid increase in the metallicity and the average old stellar population observed in the bulge nowadays \citep{Barbuy+18}.

The assumed IMF is the one of \citet{Salpeter55}:
\begin{equation}
    \varphi(M) \propto M^{-2.35}.
    \label{eq:Salpeter}
\end{equation}
We choose such an IMF since it was demonstrated in previous papers that the IMF in the bulge should top heavier than the disc IMF (\citealt{M&B90,Ballero+07,C&M2011}).
The rates of the different stellar types, SNe Ia, II, Ib and Ic are the same as in the disc, whereas in the nova rate (see eq. \ref{eq:nove}) the value $\alpha$ is set to 0.024 since it depends on the IMF of choice.

\subsection{Nucleosynthesis prescriptions}
\label{sec:nucleosynthesis}
For the nucleosynthesis prescriptions we rely on the yields already selected in \citet{Vasini+22}. 
For massive stars we assume yields from \citet{Kobayashi06}, to which we add those for $^{26}$Al from \citet{WW95} adopting the whole metallicity grid provided. For nova systems we use \citet{J&H07} that report different models regarding the nova composition (CO or ONe novae). To consider the full spectrum of possibilities in the nova population we assume that 70\% are CO novae and the remaining 30\% is composed of ONe novae. In addition, in the two cases, several yields are provided depending on the initial mass of the white dwarf. Therefore, the yield that we chose is the average of those proposed. This assumption is taken from \citet{Romano&Matteucci03} and can be summarized as:
\begin{equation}
    \langle X_{novae} \rangle = 0.7\langle X_{CO} \rangle + 0.3 \langle X_{ONe} \rangle.
    \label{eq:nova_average}
\end{equation}
The resulting nova yield is 7.69 $\times$ 10$^{-4}$ M$_{\odot}$.
For SNIa we assume yields from \citet{Iwamoto+99} and for $^{26}$Al we adopt the prescriptions from \citet{NT&Y84}. Moreover, for AGBs (M $\le$ 6 M$_\odot$) we adopt the yields from \citet{karakas2010}.

In this work we test two different models that differ only for the nucleosynthesis prescriptions summarized in Table \ref{tab:models}: Model 1 consider the production of $^{26}$Al from all the sources listed (massive stars and novae with minor contribution coming from SNIa and AGBs), and Model 2 where we exclude the contribution from novae, keeping massive star, AGBs and SNIa.  

\begin{table}[]
    \centering
    \begin{tabular}{c|c c c c}
    \hline
    \rule[-4mm]{0mm}{1cm}
        Model &  Massive stars & Novae & AGBs & SNIa \\
        \hline
        \rule[-4mm]{0mm}{1cm}
        Model 1 & \color{Green}{\cmark} & \color{Green}{\cmark} &  \color{Green}{\cmark} & \color{Green}{\cmark} \\
        \hline
        \rule[-4mm]{0mm}{1cm}
        Model 2 & \color{Green}{\cmark} & \color{Red}{\xmark} & \color{Green}{\cmark} &  \color{Green}{\cmark} \\
        \hline
    \end{tabular}

    \caption{Descriptions of the nucleosynthesis of the two models tested. The checkmarks mark the $^{26}$Al sources included in the models. For the reference to the yields adopted for each astrophysical source see Sec. \ref{sec:nucleosynthesis}.}
    \label{tab:models}
\end{table}

\section{The 2D chemical evolution model}
\label{sec:2D_model}

To extend our model from 1D to 2D, we include the azimuthal dimension by adopting the same mechanism proposed by \citet{Spitoni+19,Spitoni+23}, where SFR perturbation are implemented in order to reproduce the spiral arm pattern observed in the MW. In this Section we show how the spiral patterns are modeled and then how they are introduced in the SFR.

\subsection{Multiple spiral patterns of Spitoni+23} 
\label{sec:model_multi}
 \citet{Spitoni+23}  presents the chemical evolution of several elements in the Galactic disc, considering the effects of multiple spiral patterns on the evolution of elements synthesized over different time scales, such as oxygen, iron, europium, and barium.  
The presence of multiple spiral patterns was already predicted  by N-body simulations 
\citep{quillen2011,minchev2012,sellwood2014}
 and shown by  observations \citep{meidt2009}.

The expression for the time 
evolution  of the density perturbation, created by multiple pattern spiral arms is: 
\begin{equation}
\Sigma_{MS}(R,\phi,t)= \chi(R,t) 
\sum_{j=1}^{N} M_j(\gamma_j)\cdot \mathds{1}{\left[R_{j,\,\rm min},\  R_{j,\, \rm max}\right]},
  \label{MS_equation}
\end{equation}
where the  quantity $\chi(R,t)$ if computed at the present time $t_G$, represents the present-day amplitude of the spiral density and  can be expressed as:
\begin{equation}
\chi(R,t_G)=\Sigma_{S,0} e^{-\frac{R-R_0}{R_{S}}}.
\end{equation}
In the previous expression,  $R_S$ is  the radial scale-length of the drop-off in
density amplitude of the arms,  $\Sigma_{0}$ is the surface arm
density at fiducial radius $R_0$ fixed to values of $\Sigma_{S,0}=20 \mbox{  M}_{\odot} \mbox{  pc}^{-2}$, $R_0=8$ kpc,  and $R_S=7$
kpc as in \citet{Spitoni+23} and \citet{cox2002}.
In eq. \ref{MS_equation},  $N$ is the total number of spiral clumps and  the $M_{j} \,(\gamma_j)$ term  is the modulation function for concentrated spiral arms of \citet{cox2002}
defined for the $j^{th}$ spiral mode clump  associated with the angular velocity $\Omega_{s,j}$ and can be   expressed  as follows:
\begin{equation}
 M_{j} \,(\gamma_j)= \left(\frac {8}{3 \pi} \cos(\gamma_j)+\frac {1}{2} \cos(2\gamma_j) +\frac{8}{15 \pi} \cos(3\gamma_j)      \right),
  \label{MGAMMA}
\end{equation}
where $\gamma_j$ stands for
\begin{equation}
  \gamma_j(R,\phi,t)= m_j\left[\phi +\Omega_{s,j} \cdot t -\phi_p(R_0) -\frac{\ln(R/R_0)}{\tan(\alpha)} \right].
  \label{gamma_ref}
\end{equation}
In eq. \ref{gamma_ref}, $m_j$  refers to the multiplicity (e.g.  the number of spiral arms) associated to  the $j^{th}$ clump,  $\alpha$ is
the pitch angle \footnote{In this model all the spiral arms have the same pitch angle $\alpha$. }, $\Omega_{s,j}$ is the  angular
velocity of the pattern,   $\phi_p(R_0)$ is the
coordinate  $\phi$  computed at $t$=0 Gyr and  $R_0$.

Finally, in eq. \ref{MS_equation}, the value of the indicator function $\mathds{1}$ delimits the radial extension of the  considered spiral arm mode enclosed between the Galactocentric distances $R_{j,\,\rm min}$  and $R_{j,\,\rm max}$: it is one if the argument is within the radial interval and zero otherwise.

\begin{figure}
    \centering
    \includegraphics[scale=0.5]{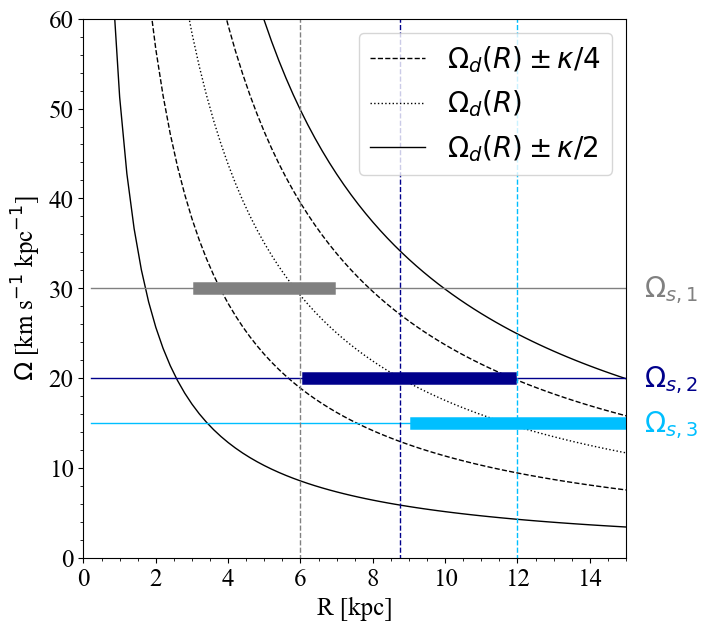}
    \caption{Spiral pattern speeds $\Omega_{s,1}(R)$, $\Omega_{s,2}(R)$ and $\Omega_{s,3}(R)$  of the  multiple spiral  modes moving at different pattern speeds of Model A in \citet{Spitoni+23} are indicated by the three coloured horizontal lines. 
  Inner and outer spiral structures (moving with the above-mentioned  pattern speeds) are also indicated  by the  thicker grey, blue and green light-blue, respectively.
    The disc angular velocity    $\Omega_d(R)$  computed by \citet{roca2014}  is indicated with the dotted line.
The 2:1 and 4:1  outer and inner Lindblad resonances (OLR and ILR) occur along the solid and dashed black curves, respectively. Resonances have been computed as $\Omega_{p2}(R) = \Omega_d(R) \pm  \kappa/2$ and $\Omega_{p4}(R) = \Omega_d(R) \pm  \kappa/4$, respectively where $ \kappa$ is the local radial epicyclic frequency. The long vertical dashed lines show the positions of the co-rotation radii assuming the three different $\Omega_S$ values.}
    \label{pattern}
\end{figure}

\begin{figure*}[h!]
    \centering
    \includegraphics[scale=0.26]{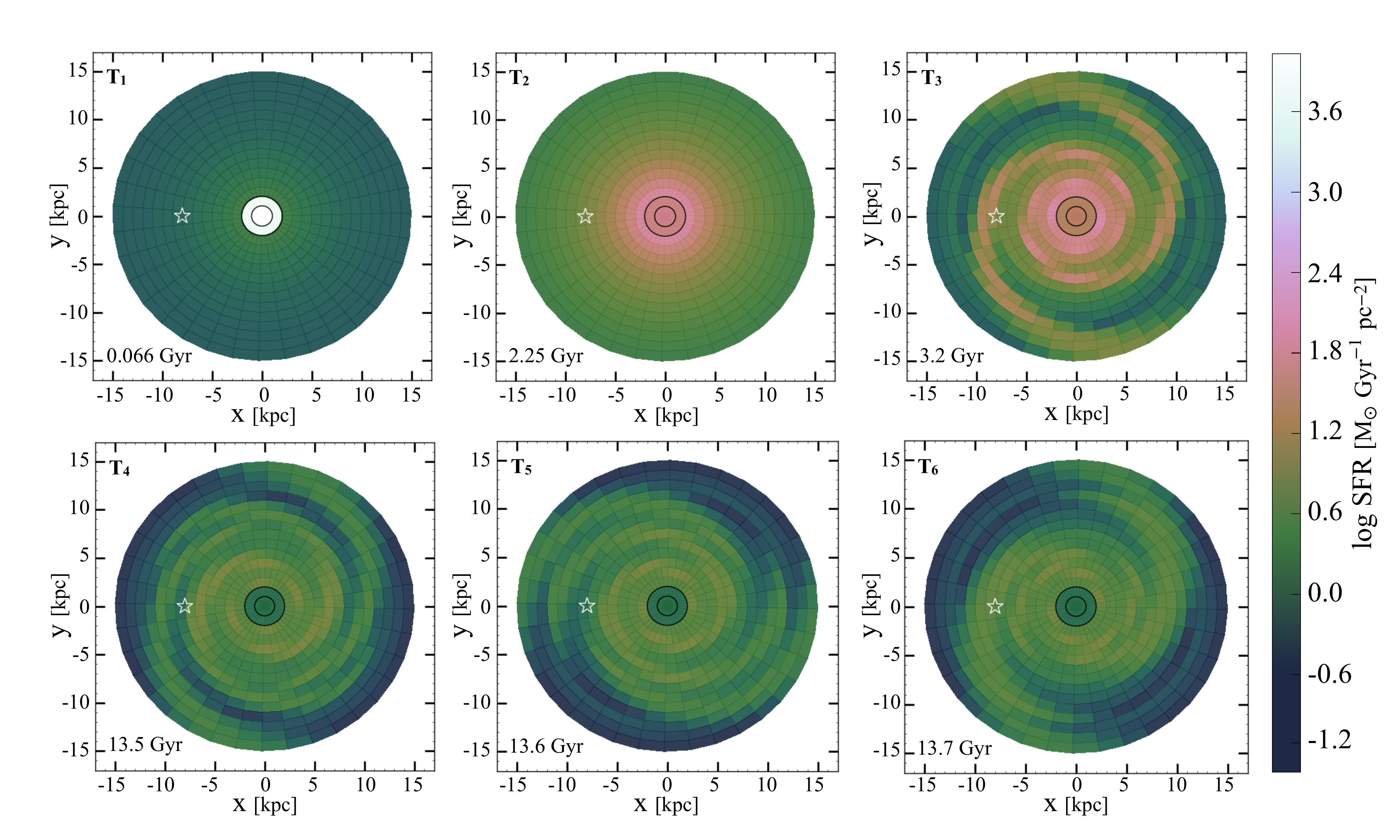}
    \caption{2D SFR from the bulge to the outskirts of the MW (assumed in this paper to be 15kpc). Here we report six snapshots at different time steps during the MW evolution: T$_1$=0.066 Gyr (the peak of the bulge SFR), T$_2$=2.25 Gyr (the peak of the disk SFR), T$_3$=3.2 Gyr (rigth after the onset of the spiral arms), two recent times, T$_4$=13.5 Gyr and T$_5$=13.6 Gyr, and the present day time, assumed to be T$_6$=13.7 Gyr. The position of the sun is represented by the white-edge star.}
    \label{fig:6panel_SFR}
\end{figure*}

 In Fig. \ref{pattern}, we show the pattern speeds $\Omega_{s,j}(R)$ of the spiral arms of Model A in \citet{Spitoni+23}, which we adopt in this work. The spiral structure is composed of three segments with multiplicity $m=2$, each moving at different pattern speeds $\Omega_{s,j}(R)$. The $j^{\text{th}}$ spiral structure is confined to the region $ R \in \left[R_{j,,\text{min}},\ R_{j,, \text{max}}\right]$.
The velocity of the central spiral structure is fixed at $\Omega_{s,2} = 20 \ \text{km s}^{-1} \ \text{kpc}^{-1}$, consistent with the \citet{roca2014} model. A similar value was first estimated from moving groups in the U-V plane by \citet[][$\Omega = 18.1 \pm 0.8 \ \text{km s}^{-1} \ \text{kpc}^{-1}$]{quillen2005}. The innermost and outermost segments rotate at velocities of $\Omega_{s,1} = 30 \ \text{km s}^{-1} \ \text{kpc}^{-1}$ and $\Omega_{s,3} = 15 \ \text{km s}^{-1} \ \text{kpc}^{-1}$, respectively.

It is interesting to note that one of the co-rotational radii is located at 8.75 kpc, which is close to the solar Galactocentric distance $(R,Z)_{\odot} = (8.249, 0.0208) \ \text{kpc}$ \citep{gravity2021,bennett2019}.

\subsection{SFR and perturbations}
\label{sec:SFR_perturbations}

We first present the assumptions adopted for the SFR and for the spiral arms in the disc ($R>2$ kpc), since the other chemical evolution prescriptions depend on those.

Our starting point is a 1D SFR depending only on the Galactocentric distance \textit{R} and following a Schmidt-Kennicutt law \citep{Kennicutt98}:
\begin{equation}
    \psi(R,t) \propto \nu(R) \Sigma_{gas}^{k}(R,t),
\end{equation}
where $\nu$ and $\Sigma_{gas}$, the star formation efficiency and the surface gas mass density respectively, both depend on the Galactocentric distance, whereas $k=1.5$ everywhere in the disc.

This relation is still mono-dimensional, meaning that we have not introduced a dependency on the azimuthal angle yet.

We  define the  adimensional perturbation  $\delta_{MS}$ as:
 \begin{equation}
    \delta_{MS}(R,\phi,t) \equiv \frac{\Sigma_D(R,t)+ \Sigma_{MS}(R,\phi,t)
      }{\Sigma_D(R,t)},
  \label{delta2}
   \end{equation}
 where $\Sigma_D$ is the
total surface mass density.
Imposing as in \citet{Spitoni+23} that the   ratio $\chi(R,t)/\Sigma_D(R,t)$ is constant in time, the  adimensional perturbation  $\delta_{MS}$ defined above becomes:
  \begin{equation}
    \delta_{MS}(R,\phi,t) =1 + \frac{
      \chi(R,t_G)}{\Sigma_D(R,t_G)} \sum_{j=1}^{N} M_j(\gamma_j)\cdot \mathds{1}{\left[R_{j,\,\rm min},\  R_{j,\, \rm max}\right]},
  \label{delta22}
   \end{equation}
%This dimensional quantity has the important feature that its azimuthal average value at a fixed Galactocentric distance $R$ and time $t$ is, since for each spiral clump we have that  $<M_j(\gamma_j)>_{\phi}=0$ \citep{cox2002}.
Finally, the new SFR in presence of the effect of spiral arms is computed as:
\begin{equation}
    \psi(R,t,\phi) \propto \nu(R) \Sigma_{gas}^{k}(R,t) \cdot  \delta_{MS}^k(R,\phi,t).
\end{equation}

\begin{figure*}
    \centering
    \includegraphics[scale=0.2]{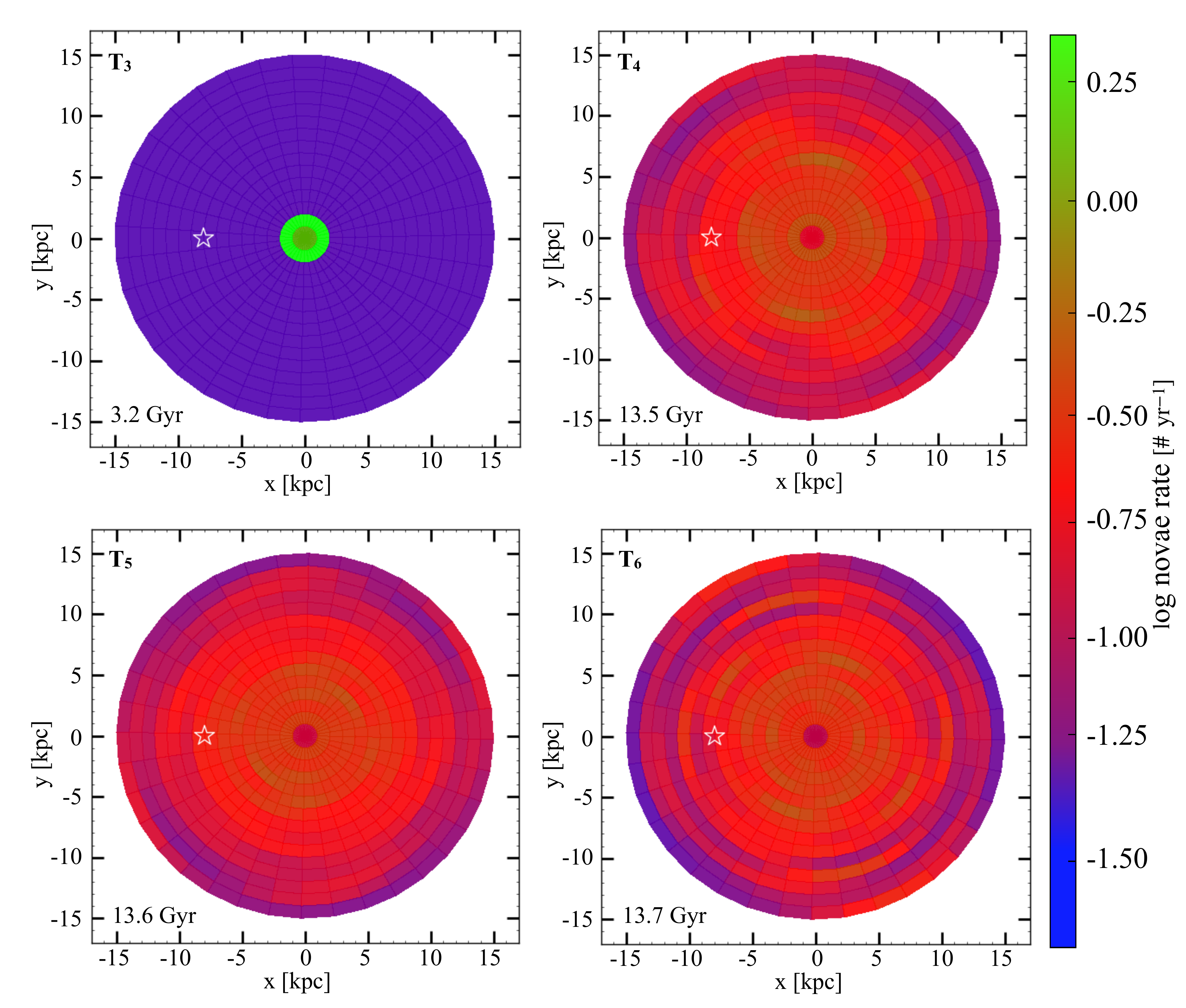}
    \caption{2D map of nova outburst rate in the MW at four different times, T$_3$, T$_4$, T$_5$ and T$_6$.}
    \label{fig:novae}
\end{figure*}

We apply the perturbation after 3 Gyr of evolution and then we let the MW evolve up to 13.7 Gyr, the assumed age of the Universe. We make this choice in order to mimic what is done in \citet{Spitoni+19}. They simulate the 2D evolution of the thin disc only, assuming a one-infall episode lasting for 11 Gyr. In our case, where we model both the thin and the thick discs in the framework of a two-infall scenario for 13.7 Gyr, we delay the rising of the perturbation by 3 Gyr relative to the beginning of the star formation, so that also in our case the evolution of the perturbation lasts for roughly 11 Gyr.

%{\bf FRASE SEGUENTE VA RIVISTA, NON CHIARA} where the evolution of the Milky Way with perturbed SFR was carried out at around 11 Gyr. Since we assume that the current estimate of the age of the Universe is 13.7 Gyr, applying the pertu

We show our perturbed SFR in Fig. \ref{fig:6panel_SFR}. We represent both the bulge (R$\le$ 2 kpc) and the disc at six different times using the same color coding so as to appreciate the SFR variations in time and space. In the top row, starting from the left panel the snapshots shown are those corresponding to  the peak of the bulge SFR (T$_1$=0.066 Gyr), the peak of the disc SFR (T$_2$=2.25 Gyr), and the rising of the SFR perturbations (T$_3$=3.2 Gyr). Regarding the bottom panels, when dealing with short-lived radioisotopes as $^{26}$Al the observations can recover only the recent history which is related to the recent SFR, therefore we show three recent times (T$_4$=13.5 Gyr, T$_5$=13.6 Gyr and T$_6$=13.7 Gyr) to hopefully trace their SFR when analyzing the $^{26}$Al distribution. The white edge star shows the locations of the the solar neighbourhood. 

In the T$_1$ panel we show that when the bulge SFR reaches a maximum,  the disc SFR is still very low, roughly four order of magnitudes lower. Then, at T$_2$, the bulge SFR has already decreased since it is characterized by an extremely efficient and short burst whereas the disc SFR shows a spike, whose value is smaller and smaller as we go towards the outskirts. At T$_3$ the perturbations arise and are clearly visible in the disc since this is the moment when they are at their peak. After this moment the SFR perturbations will monotonically decrease. 
In the three bottom panels (T$_4$, T$_5$ and T$_6$) the perturbations are still evident, although quite weaker, and the different segments that compose the single arms are not aligned due to their different rotational velocities. The result is that the arms look broadened as it is particularly evident in the last panel.

\begin{figure*}[h]
    \centering
    \includegraphics[scale=0.26]{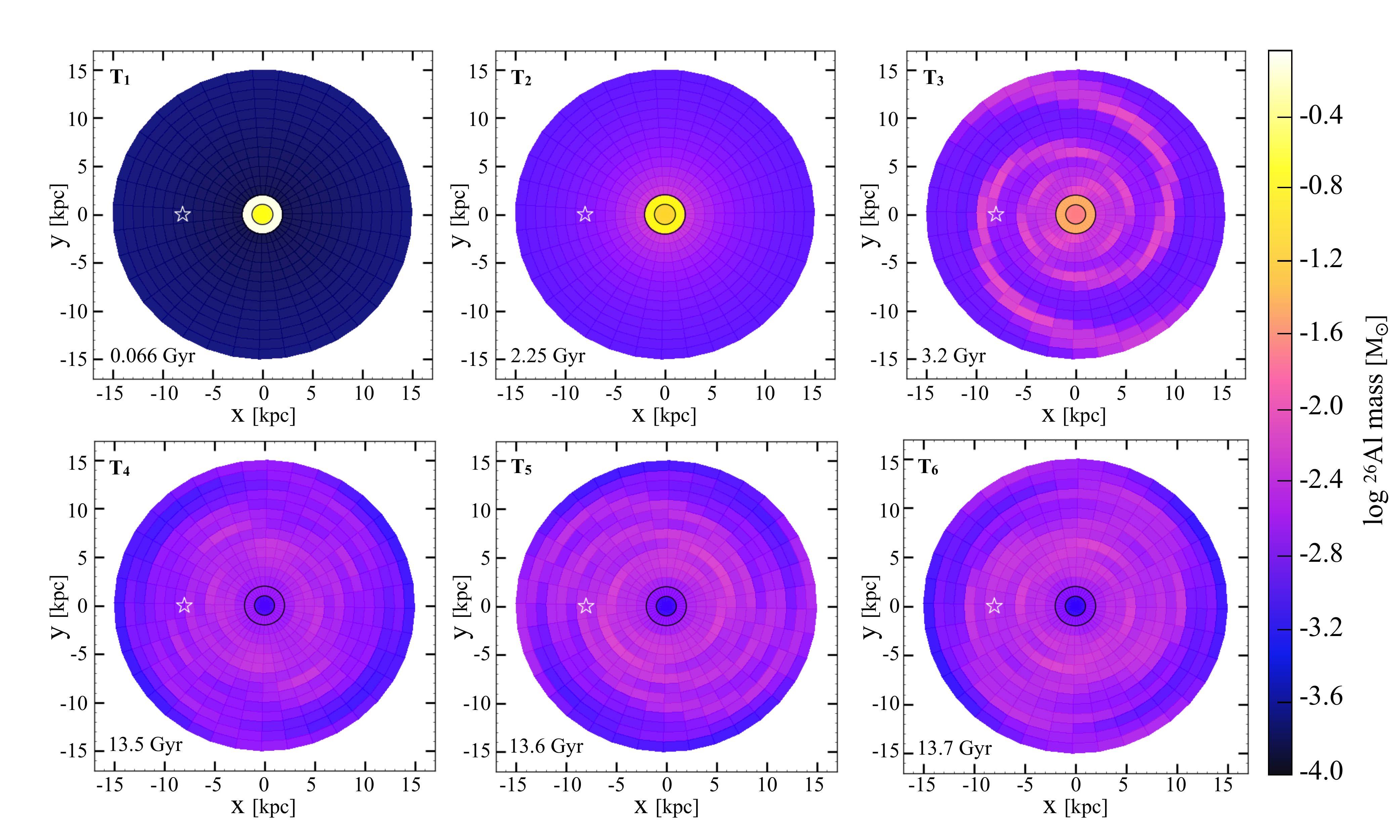}
    \caption{In this panel we show the 2D distribution of $^{26}$Al mass at the six time steps where we have previously analyzed the SFR. In this case we accounted as $^{26}$Al producers both massive stars and novae. The bulge is not represented. Each cell is color coded according to the $^{26}$Al mass contained at that time step and as in Fig. \ref{fig:6panel_SFR} the sun is indicated by the white-edge star.}
    \label{fig:Al_T1}
\end{figure*}

\section{Results}
\label{sec:result_all}
 In this Section we show the results regarding the nova and $^{26}$Al 2D distribution in the MW, from the bulge up to the outermost region of the Galaxy.
%{\bf TUTTA LA PARTE RISULTATI E? DI DIFFICILE LETTURA; E' PIENA DI RIPETIZIONI: NON METTO PARTICOLARI CORREZIONI PERCHE' DOVRESTIRISCRIVERLA PIU' SNELLA. E' MOLTO IMPORTANTE FARLA PIU SNELLA E PIU'CHIARA
\subsection{Nova systems}
\label{sec:result_novae}

In this Section we show the results related to the nova outbursts in the MW. 

In Fig. \ref{fig:novae} we report the 2D map of the novae for four of the six times shown in Fig. \ref{fig:6panel_SFR}. The four panels refer to times T$_3$, T$_4$, T$_5$ and T$_6$ as defined in Sec. \ref{sec:SFR_perturbations}.
We do not show the snapshots at T$_1$ and T$_2$ since at such early epochs novae have not started to explode yet.
%leaving out the first two since at early epochs the novae have not yet started exploding due to their delay time distribution.
The maps are color coded according to the logarithm of the nova outburst rate, and the solar ring at 8 kpc is marked by the white-edge star in the left side of each panel. 

The difference between the early epoch (T$_3$) and the recent ones (T$_4$, T$_5$ and T$_6$) is evident. The reason, as anticipated, lies in the delay time distribution of the novae.

Because of this delay, novae do not trace the SFR, and no correspondence can be found between the 2D SFR map and the 2D nova map computed at the same epoch. Therefore, at T$_3$ the nova distribution is still uniform, even though the SFR is already perturbed, and at late epochs novae do not show the spiral arm pattern which is shown by the SFR.
%The SFR in the disc peaks at T$_2$, therefore at T$_3$ very few novae started exploding. Moreover, even though the SFR was already perturbed the novae are not. This is the evidence for the late effect of the SFR on the nova outbursts. 

The same time delay affects the bulge novae but the result in this region is slightly different, since the SFR in the bulge is more efficient. Hence, the majority of the bulge novae is produced earlier, at T$_1$ when the bulge SFR peaks, and therefore they explode earlier. Hence, at T$_3$ the nova outburst rate in the bulge is already high.

%The scenario is different for the bulge, where the SFR peaks at T$_1$, anticipating the novae contribution that results in a spike of the novae outbursts at T$_3$. 
%The recent time panels appear very different from the T$_3$ panel. Regarding the disc, the nova rate is much larger than the T$_3$ one and is now inhomogeneous, since the effect of the perturbation have kicked in. The bulge is slightly poorer since the novae peaked previously. The inhomogeneities are evident but are not reproducing the spiral arm pattern since the delay time distribution delays different systems in different ways, hence the novae exploding at a specific time were not produced all at the same epoch. Each nova then traces the SFR at its own formation epoch, therefore at the present time the net effect is the loss of distinct spiral pattern.

\subsection{$^{26}$Al mass 2D distribution}
\label{sec:result_al_map}

\begin{figure*}
    \centering
    \includegraphics[scale=0.26]{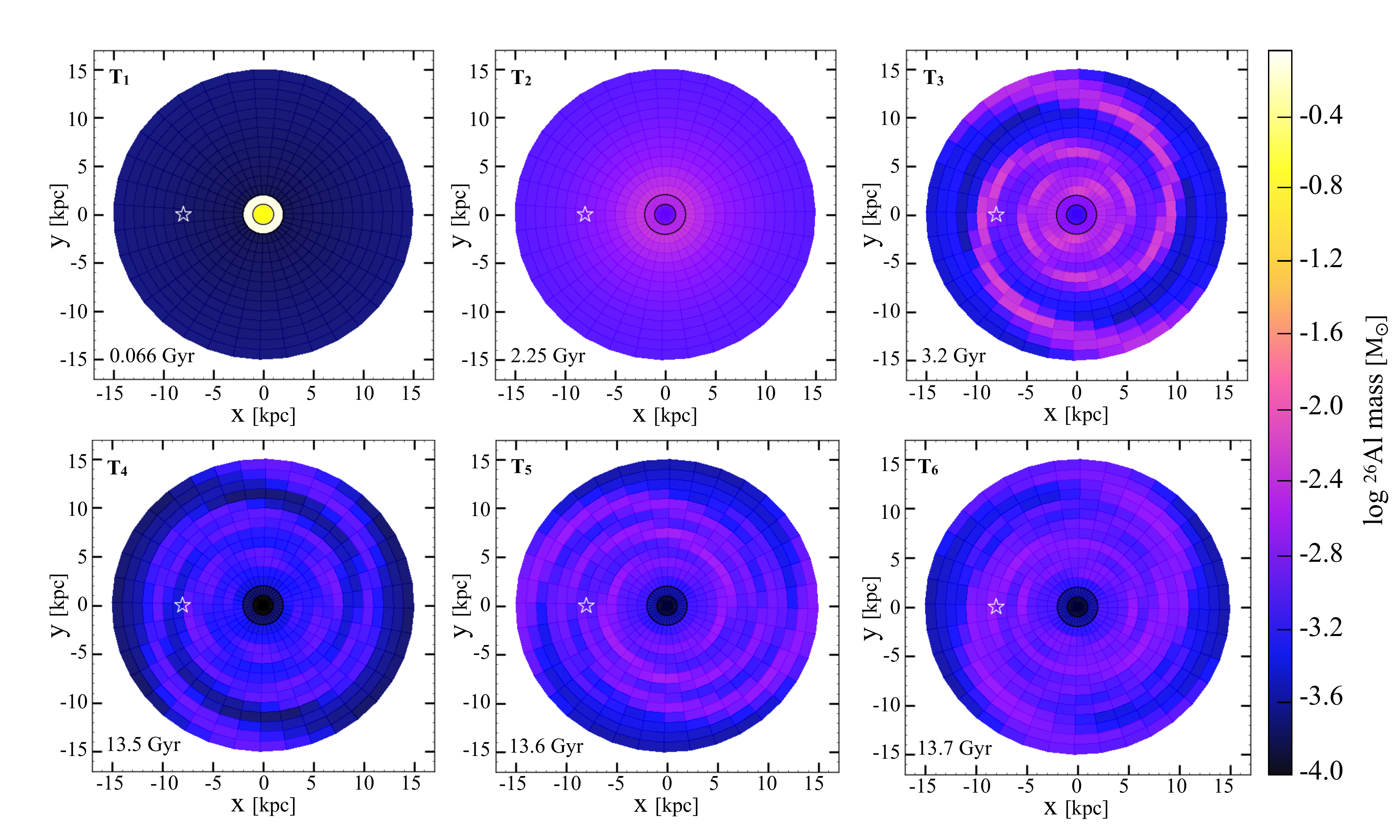}
    \caption{Same as Fig. \ref{fig:Al_T1} but without any contribution from nova systems.}
    \label{fig:Al_SN1}
\end{figure*}

In this Section, we show the results related to the 2D distribution of $^{26}$Al in the MW within 15 kpc from the galactic center. 

In Fig. \ref{fig:Al_T1} we report the results from Model 1, where massive stars and nova systems both contribute to the synthesis of $^{26}$Al. In this figure we show the snapshots at the same six times as in Fig. \ref{fig:6panel_SFR}. The plots are color coded according to the logarithm of the $^{26}$Al mass in each cell. The indicative location of the solar neighbourhood is suggested by a white-edge star in the left side of each snapshot. 
\newline

A comparison between Fig. \ref{fig:Al_T1} and the SFR in Fig.\ref{fig:6panel_SFR} is quite interesting.
In Fig. \ref{fig:Al_T1} the first three panels (T$_1$, T$_2$ and T$_3$) reproduce perfectly the SFR pattern:
\begin{itemize}
    \item at T$_1$=0.066 Gyr the only contribution to $^{26}$Al in the MW comes from the bulge (R < 2 kpc) where the SFR is high, whereas $^{26}$Al in the disc is still very low (as it is the SFR);
    \item at T$_2$=2.25 Gyr the SFR in the bulge has already reached a maximum and decreased and the $^{26}$Al shown in the plot has originated from the nova explosions. 
    On the other hand, the SFR in the disc reached the peak, thus causing an increment in the amount of $^{26}$Al produced. Nevertheless, the peak of the $^{26}$Al production in the disc will occur slightly later, due to the delayed contribution from nova systems;
    %The reason for this delay is related to the timescales of the two main $^{26}$Al production sites, massive stars and novae, and to the intensity of the SFR. If the SFR is extremely high and narrow in time, as in the case of the bulge, the $^{26}$Al will show a prominent spike corresponding to the SFR peak and will rapidly decrease. The nova contribution will kick in later (when $^{26}$Al from massive stars have already decayed) and will produce a second increase in the $^{26}$Al mass, which is smaller than the first one. Therefore, for burst-like SFR the burst of the SFR corresponds to the $^{26}$Al peak. If the SFR follows a smoother behaviour, such as in the disc, the peak of $^{26}$Al occurs at later times, between the SFR peak and the nova peak;
    \item At T$_3$=3.2 Gyr, the SFR perturbations have just kicked in and the spiral pattern is accurately reproduced by the $^{26}$Al 2D distribution. 
\end{itemize}

 These first three panels perfectly mimic the SFR behaviour because after few billion years of evolution, the only significant contribution to the $^{26}$Al comes from the massive stars. Novae contribution will arise at later times, therefore in the first stages of the evolution it is still negligible.

On the other hand, at T$_4$, T$_5$ and T$_6$ the scenario is different: in every snapshot the SFR pattern is reproduced with less accuracy and the spiral pattern is recognisable only in the outermost rings.
%By comparing these three snapshots with the corresponding SFR it is evident that . 
%In the innermost annuli the only evidence left of a spiral pattern SFR is given by the weak $^{26}$Al peaks, spread among several neighbouring cells. 
Regarding the present time bulge, the massive stars are almost totally absent and the $^{26}$Al predicted there comes exclusively from recent nova outbursts.
%The reason behind this, is that novae are not negligible anymore due to their delay time distribution and the net result is a smoothing of the oscillations in the $^{26}$Al mass.
\newline

In Fig. \ref{fig:Al_SN1}, where the results from Model 2 are plotted, the scenario is rather different, since novae have been excluded from the $^{26}$Al nucleosynthesis. The snapshots are taken at the same times, the color coding is the same and the white-edge star still represent the indicative position of the solar neighbourhood.

The first three panels resemble Fig. \ref{fig:Al_T1}. The only difference is the overall depletion of $^{26}$Al mass over the entire Galaxy due to the removal of the nova systems. The bulge is much less enriched in $^{26}$Al (see T$_2$ panel) with respect to Fig. \ref{fig:Al_T1}, since the only contribution in this region was coming from nova systems, that are now excluded. 
%and more than two Gyr after the burst of SFR in the bulge very few massive stars are left. 
On the other side, the three recent snapshots (T$_4$, T$_5$ and T$_6$) show a rather good tracing of the SFR since no smoothing from novae is involved. The bulge looks extremely poor in $^{26}$Al since the massive stars are completely absent. The only recent production in the region R < 2 kpc is that coming from AGBs and SNIa.

\subsection{$^{26}$Al mass azimuthal oscillations}

To quantify how much the novae contribute to the scenario we present Fig. \ref{fig:oscillazioni}:
\begin{itemize}
    \item panel a: we show the SFR at the present time, in the solar ring as a function of the azimuth. The vertical dashed lines mark the peaks of the SFR and are reported in all the panels below. The oscillations are symmetric as was already shown in Fig. \ref{fig:6panel_SFR};
    \item panel b: we represent the $^{26}$Al mass for Model 1 and Model 2 (pink and green lines respectively). The vertical dashed lines refer to the peaks of the SFR. 
    The oscillations have the same amplitude in the two cases, but, as shown in Fig. \ref{fig:Al_T1} and \ref{fig:Al_SN1}, they are far smoother in Model 1 where the spiral pattern at the present time is almost lost. The relative enhancement of the peaks with respect to the minima: for Model 1 the peak is the 28\% higher than the minima, whereas in the case of Model 2 the peak is the 150\% higher than the minima. The peaks in Model 2 stand out against the background by a factor of $\sim$5 more than in Model 1, hence the loss of the spiral pattern;
    \item panel c: we plot the difference between the $^{26}$Al produced by Model 1 and Model 2. Here we show clearly that the mass of $^{26}$Al produced by novae is dependent  on the azimuth but not on the SFR. This value oscillates between 0.004 and 0.006 M$_{\odot}$ and is largest in the minima of the SFR, demonstrating why novae blunt the spiral pattern;
    \item panel d: here we show the nova distribution in the solar ring, that resembles that of the $^{26}$Al in the previous panel.
\end{itemize}
\begin{figure*}
    \centering
    \includegraphics[scale=0.8]{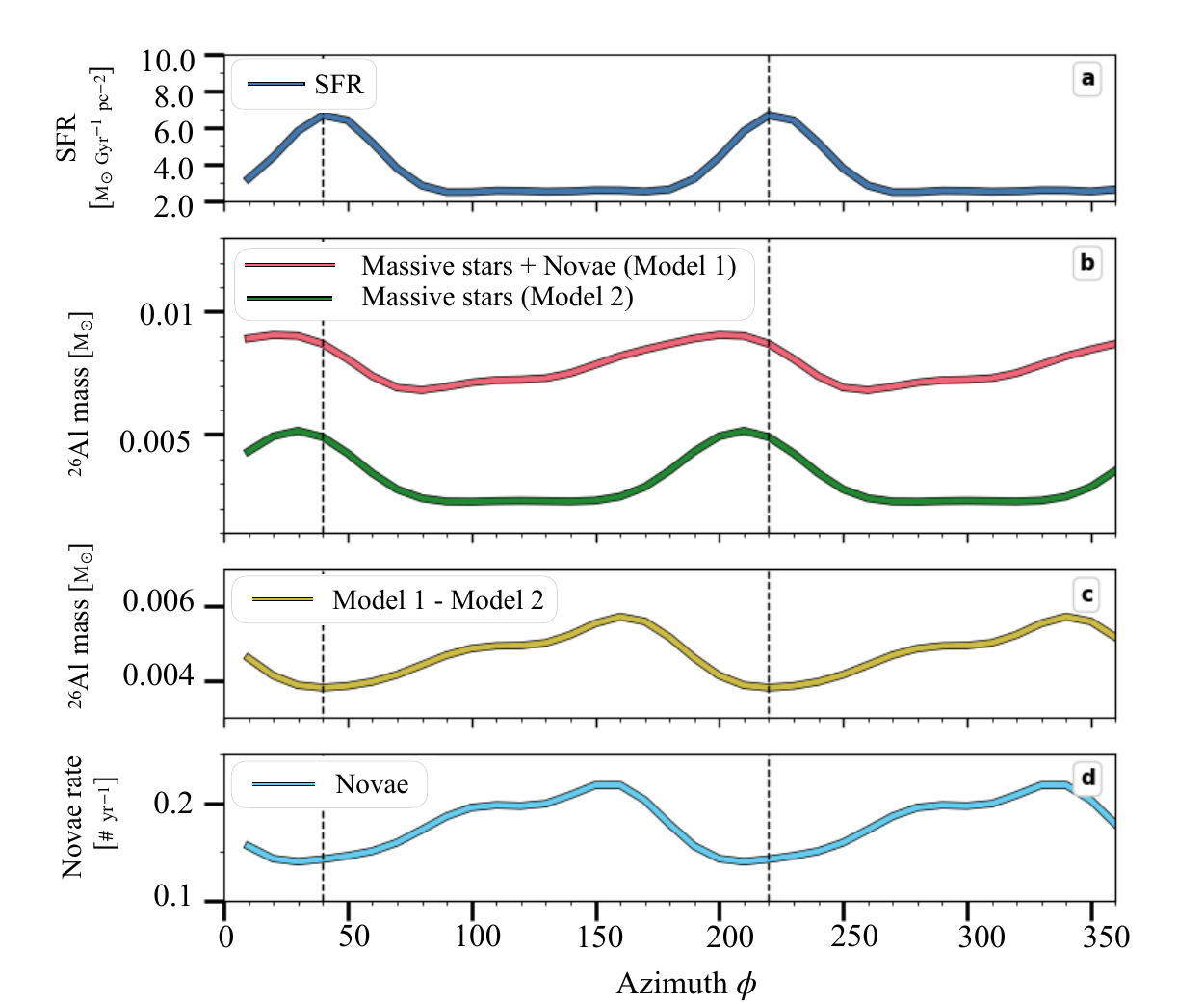}
    \caption{\textit{Panel a}: SFR as a function of the azimuth at the present time and the solar neighbourhood. The black dashed lines, in this plot and in those below, mark the peaks of the SFR. \textit{Panel b}: $^{26}$Al distribution in the solar ring and present time. In pink, $^{26}$Al predicted by Model 1 and in green that from Model 2. \textit{Panel c}: difference between the $^{26}$Al production in the two cases. \textit{Panel d}: novae distribution in the solar ring at the present time.}
    \label{fig:oscillazioni}
\end{figure*}

\begin{table*}
\centering
\begin{tabular}{c|c|c|c|c|c|c}
\hline
%\setlength{\extrarowheigth}{100pt}
%\arraystretch
\rule[-4mm]{0mm}{1cm}
 \textbf{Model} & \textbf{0 - 1 kpc} & \textbf{1 - 2 kpc} & \textbf{2 - 3 kpc} & \textbf{3 - 4 kpc} & \textbf{4 - 5 kpc} & \textbf{Total (0 - 5 kpc)} \\
 \hline 
 \rule[-4mm]{0mm}{1cm}
    \textbf{Model 1} & 0.056 M$_{\odot}$ & 0.150 M$_{\odot}$ & 0.229 M$_{\odot}$ & 0.283 M$_{\odot}$ & 0.312 M$_{\odot}$ & 1.028 M$_{\odot}$ \\
 \hline 
 \rule[-4mm]{0mm}{1cm}
    \textbf{Model 2} & 0.006 M$_{\odot}$ & 0.014 M$_{\odot}$ & 0.057 M$_{\odot}$ & 0.083 M$_{\odot}$ & 0.105 M$_{\odot}$ & 0.265 M$_{\odot}$ \\
\hline
\end{tabular}
\caption{$^{26}$Al integrated over the five innermost annuli (0 - 5 kpc) at the present time for both models. Column 2-6 report the $^{26}$Al for every single ring (each 1 kpc wide) and column 7 show the total $^{26}$Al within 5 kpc.}
\label{tab:al_tabella}
\end{table*}

%where  (top panel), , the  (third panel) and the nova outburst rate (bottom panel). All the quantities are represented at the present time, in the solar ring as a function of the azimuth. We marked with vertical dashed black lines the peaks of the SFR and reported them to the panels below. In the second panel from the top the green line is model M2, produced by excluding novae from the nucleosynthesis. Therefore, the oscillations mirror exactly the SFR. 
%Even though they die quickly after their birth, we cannot assume that they all instantly die as it is usually done in chemical evolution. This is because, for the last 200 Myr of the simulation, we set the timesteps to be 0.5 Myr long, shorter than the lifetime of a fraction of massive stars, hence there is a small delay in the massive star explosion, so that the $^{26}$Al oscillations actually trace the SFR of a few Myr before the current one. The same displacement is evident for model M2 since massive stars are the same. 

%The increase in $^{26}$Al contribution due to the novae is evident in model M1 (pink line) and, most important, almost homogeneous (novae don't depend strongly on the SFR). 

\subsection{$^{26}$Al: predictions versus observations}

Regarding the comparison with observational data in Table \ref{tab:al_tabella} we report the values of $^{26}$Al within 5 kpc from the Galactic center. Columns from 2-6 report the values in the 1 kpc wide annuli and their sum is listed in column 7. 
Up to now, the observations performed with COMPTEL and INTEGRAL agreed on the amount of $^{26}$Al present in the MW. 
Adopting the COMPTEL data \citealt{Prantzos&Diehl96}
estimated an $^{26}$Al mass in the range 1.5 - 2.0 M$_{\odot}$ within 5 kpc from the Galactic centre (see also \citealt{Diehl+95}). Later, \citealt{Diehl+16} estimated 2.0 $\pm$ 0.3 M$_{\odot}$ of $^{26}$Al in the same region after the INTEGRAL data were released.

Within 5 kpc from the Galactic center our Model 1 predicts 1.028 M$_{\odot}$ of $^{26}$Al, whereas Model 2 predicts 0.265 M$_{\odot}$. In neither of the two cases we are able to recover the exact amount of the $^{26}$Al observed, but a potential solution was proposed by \citet{dV&I2020}. The authors report that the nucleosynthesis of the novae in the bulge could be significantly different from that of the disc novae. The bulge novae could actually eject up to a factor of ten more matter than the disc novae, therefore increasing of the same factor the $^{26}$Al produced. This can be justified by considering that bulge novae exploding now come from very low mass stars that need to accrete a thicker envelope to explode, thus ejecting more matter.} By including this additional factor in the calculations, we can reproduce up to 2.882 M$_{\odot}$ of $^{26}$Al with Model 1, therefore with a slightly smaller factor we can reproduce the $\sim$2 M$_{\odot}$ observed.
%As reported in the table neither of the two models reproduces this quantity even though there are no other additional known production sites to consider and the yields had already been constrained by \citet{Vasini+22}. The solution may come from \citet{dV&I2020} were the authors report that the nucleosynthesis of the novae in the bulge could be significantly different from that of the disc novae. The bulge novae could actually eject up to a factor of ten more matter than the disc novae, therefore increasing of the same factor the $^{26}$Al produced. By including this additional factor in the calculations, model M1 predicts 2.882 M$_{\odot}$ of $^{26}$Al, compatible with the observational uncertainties. This correction cannot be included in model M2, due to the absence of $^{26}$Al production from nova systems, making this model even less plausible.

\subsection{Galactic observational constraints}
\label{sec:constraints}

\begin{table}[h!]
    \centering
    \begin{tabular}{c|c c}
         \hline
         \rule[-4mm]{0mm}{1cm}
         & Observations & Predictions \\
         \hline
         %\rule[-4mm]{0mm}{1cm}
         SFR [M$_{\odot}$ yr$^{-1}$ ] & 0.65 - 3 (1) & 2.24\\
         %\hline
         %\rule[-4mm]{0mm}{1cm}
         Gas mass [10$^9$ M$_\odot$] & 8.1 - 4.5 (2) & 6.4\\
         %\hline
         %\rule[-4mm]{0mm}{1cm}
         Stellar mass [10$^{10}$ M$_\odot$] & 3. - 4. (3) & 3.7 \\
         %\hline
         %\rule[-4mm]{0mm}{1cm}
         Infall rate [M$_\odot$ yr$^{-1}$] & 0.6 - 1.6 (4) & 0.65 \\
         %\hline
         %\rule[-4mm]{0mm}{1cm}
         Type II SNe [events century$^{-1}$] & 0.4 - 2.0 (5) & 0.88 \\
         %\hline
         %\rule[-4mm]{0mm}{1cm}
         Type Ia SNe [events century$^{-1}$] & 0.1 - 0.5 (5) & 0.2 \\
         \hline
    \end{tabular}
    \caption{MW disc (R>2 kpc) constraints: in column 2 the observed quantities, in column 3 the predictions from our models. Models 1 and 2 predict the same observables since they differ only on the nucleosynthesis side. \textbf{References:} (1) \citet{Prantzos+11}, \citet{Chomiuk&Povich11}, (2) \citet{KPA15}, (3) \citet{Flynn+06}, (4) \citet{Marasco+12}, \citet{L&H11}, (5) \citet{cappellaro1997}.}
    \label{tab:constraints}
\end{table}

With our choice of input parameters we are able to reproduce the main observational features of the MW disc (R>2 kpc). We highlight in particular that we can reproduce the current SFR, gas mass, stellar mass, infall rate and both Type Ia and Type II SNe rates.

We report in Table \ref{tab:constraints} the most important observational constraints in the Galactic disc: in column 2 we show the observed value with the reference, in column 3 the predicted value by our models. We compare our results with the observations from different authors. Regarding the SFR we refer to the values proposed by \citet{Prantzos+11} and \citet{Chomiuk&Povich11}, for the gas mass we rely on \citet{kubryk2015}, for the stellar mass we compare our results to \citet{Flynn+06}, for the infall rate we use the observations by \citet{Marasco+12} and \citet{L&H11} and for the Type II and Type Ia SN rate we refer to \citet{cappellaro1997}.
We highlight that the two models tested, Model 1 and Model 2, differ only for the nucleosynthesis, therefore the predictions of all the unrelated quantities (such as those listed here) are the same.

\section{Conclusions}
\label{sec:conclusions}
In this paper we model the temporal ad spatial evolution of $^{26}$Al in the MW by adopting a 2D chemical evolution model to perform a comparison with the already existing data and a prediction for the upcoming ones.The introduction of a second dimension in the chemical evolution model allows us to investigate how much the approximation of homogeneous mixing affects the production of $^{26}$Al, and which is the relative contribution of novae and massive stars to its mass in the MW. Moreover we can also probe how good the tracing of the SFR is when considering nova systems as $^{26}$Al producers. We assume the Galaxy to be divided into two main regions: the bulge (R $\le$ 2 kpc), that is described adopting a homogeneous 1D model, and the disc (R > 2 kpc) where we consider a 2D model, with the SFR being dependent not only on the Galactocentric distance and time but also on the azimuth. The dependency on the azimuth is introduced by applying a perturbation to the 1D SFR, resembling the rotating spiral arm pattern observed in the Galaxy. Our main focus is $^{26}$Al, a short-lived radioisotope with a $\sim$1 Myr decay time-scale, generally assumed to be  produced by massive stars. These two facts suggest that  $^{26}$Al can be a good tracer of active star formation, therefore its current location in the MW is dependent on the shape of the recent SFR. 

With this model we aim  at reproducing the observed mass of $^{26}$Al in the Galaxy to investigate the importance of nova contribution to the mass of this element, and studying the effects that novae  can produce on  $^{26}$Al as tracer of star formation.  To do that, we compare two models considering different $^{26}$Al sources: Model 1, where massive stars and nova systems are both responsible for the $^{26}$Al production, and Model 2 where nova systems are excluded. Regarding the yields for both the production sites we adopt the prescriptions of \citet{Vasini+22}. 
By means of our chemical evolution model, we are able to produce 2D maps for the mass of $^{26}$Al, the distribution of massive stars and nova systems in the MW. 
Our main conclusions can be summarised as follows:

\begin{itemize}
    \item the nova systems eject nucleosynthesis products with a large time delay due to the initial masses of the binary components and to the time necessary for a white dwarf to cool. Due to this delay, novae are not tracers of the SFR. The maps that we produce show that that the peaks of the present time nova distribution at the solar neighbourhood are located in the minima of the present time SFR;
    %{\bf  questa frase che segue non e' chiara, perche' le novae dovrebbero essere distribuite in maniera non omogenea?the 2D nova distribution starts being inhomogeneous at later times with respect to the rise of the perturbation of the SFR.}
    \item the two hypotheses about the producers of $^{26}$Al that we test, predict different distributions of the mass of this element. We highlight that when including the production by novae (Model 1), the spiral pattern at the present time is hardly visible,  at variance with the case of production from massive stars only (Model 2). In particular, in the case of $^{26}$Al from both production sites, the peaks of the $^{26}$Al distribution are $\sim$ 28\% higher that the minima, whereas in the case of $^{26}$Al from massive stars only the peaks are $\sim$ 150\% higher. Therefore, in the case without novae, the peaks of the distribution stand out from the background five times more than in the case with novae included, thus tracing better the spiral arm pattern; 
    \item we compare the $^{26}$Al predicted by the two models with the available observational data, which refer to the innermost 5 kpc of the Galaxy. We integrate the $^{26}$Al within each ring from the center up to 5 kpc and we obtain 1.028 M$_{\odot}$ if novae are included (Model 1) and 0.265 if novae are excluded (Model 2). Therefore, the relative contribution of novae to the total amount of $^{26}$Al is $\sim$75\% and the other $\sim$25\% comes from massive stars. The observations report $\sim$ 2 M$_{\odot}$ of $^{26}$Al, hence both our models underestimate the mass of $^{26}$Al.
    The missing $^{26}$Al mass can be recovered if we consider that bulge novae can eject up to 10 times more matter than disc novae. Considering a factor of ten, our Model 1 predicts 2.882 M$_{\odot}$ of $^{26}$Al within 5 kpc from the Galactic center, therefore by adopting a factor of $\sim$6 we can very well reproduce the observations.
\end{itemize}

In conclusion, since the tracing of the SFR by $^{26}$Al is lowered by a factor of $\sim$5, if novae are considered, we cannot say that this element is a pure tracer of the SFR. Therefore, other SLRs not produced by novae, such as $^{60}$Fe are more reliable tracers of the SFR.

From the nucleosynthesis side, our 2D chemical evolution model confirms that the nova contribution to $^{26}$Al is fundamental to reproduce the observations especially in the bulge, where they should eject a larger amount of  chemical elements as compared to those produced by the disc novae.

\section*{Acknowledgement}
  A. Vasini  and  F. Matteucci thank I.N.A.F. for the 1.05.12.06.05 Theory Grant
- Galactic archaeology with radioactive and stable nuclei.  
This research was supported by the Munich Institute for Astro-, Particle and BioPhysics (MIAPbP) which is funded by the Deutsche Forschungsgemeinschaft (DFG, German Research Foundation) under Germany´s Excellence Strategy – EXC-2094 – 390783311.
F. Matteucci thanks also support from Project PRIN MUR 2022 (code 2022ARWP9C) "Early Formation and Evolution of Bulge and HalO (EFEBHO)" (PI: M. Marconi), funded by the European Union – Next Generation EU. 
G.Cescutti thanks the support from the PRIN MUR project no. 2022X4TM3H “Cosmic POT” (PI Laura Magrini), also funded by the European Union – Next Generation EU.
E. Spitoni and G. Cescutti thank I.N.A.F. for the  
1.05.23.01.09 Large Grant - Beyond metallicity: Exploiting the full POtential of CHemical elements (EPOCH) (ref. Laura Magrini).

\bibliographystyle{aa} % style aa.bst
\bibliography{2D_arianna}

\begin{thebibliography}{66}
\expandafter\ifx\csname natexlab\endcsname\relax\def\natexlab#1{#1}\fi

\bibitem[{{Ballero} {et~al.}(2007){Ballero}, {Matteucci}, {Origlia}, \& {Rich}}]{Ballero+07}
{Ballero}, S.~K., {Matteucci}, F., {Origlia}, L., \& {Rich}, R.~M. 2007, \aap, 467, 123

\bibitem[{{Barbuy} {et~al.}(2018){Barbuy}, {Chiappini}, \& {Gerhard}}]{Barbuy+18}
{Barbuy}, B., {Chiappini}, C., \& {Gerhard}, O. 2018, \araa, 56, 223

\bibitem[{{Bennett} \& {Bovy}(2019)}]{bennett2019}
{Bennett}, M. \& {Bovy}, J. 2019, \mnras, 482, 1417

\bibitem[{{Bird} {et~al.}(2013){Bird}, {Kazantzidis}, {Weinberg}, {Guedes}, {Callegari}, {Mayer}, \& {Madau}}]{bird2013}
{Bird}, J.~C., {Kazantzidis}, S., {Weinberg}, D.~H., {et~al.} 2013, \apj, 773, 43

\bibitem[{{Brook} {et~al.}(2012){Brook}, {Stinson}, {Gibson}, {Kawata}, {House}, {Miranda}, {Macci{\`o}}, {Pilkington}, {Ro{\v{s}}kar}, {Wadsley}, \& {Quinn}}]{brook2012}
{Brook}, C.~B., {Stinson}, G.~S., {Gibson}, B.~K., {et~al.} 2012, \mnras, 426, 690

\bibitem[{{Cappellaro} \& {Turatto}(1997)}]{cappellaro1997}
{Cappellaro}, E. \& {Turatto}, M. 1997, in NATO Advanced Science Institutes (ASI) Series C, Vol. 486, NATO Advanced Science Institutes (ASI) Series C, ed. P.~{Ruiz-Lapuente}, R.~{Canal}, \& J.~{Isern}, 77

\bibitem[{{Cescutti} \& {Matteucci}(2011)}]{C&M2011}
{Cescutti}, G. \& {Matteucci}, F. 2011, \aap, 525, A126

\bibitem[{{Cescutti} \& {Molaro}(2019)}]{Cescutti&Molaro19}
{Cescutti}, G. \& {Molaro}, P. 2019, \mnras, 482, 4372

\bibitem[{{Chiappini} {et~al.}(2001){Chiappini}, {Matteucci}, \& {Romano}}]{Chiappini+2001}
{Chiappini}, C., {Matteucci}, F., \& {Romano}, D. 2001, \apj, 554, 1044

\bibitem[{{Chomiuk} \& {Povich}(2011)}]{Chomiuk&Povich11}
{Chomiuk}, L. \& {Povich}, M.~S. 2011, \aj, 142, 197

\bibitem[{{C{\^o}t{\'e}} {et~al.}(2019{\natexlab{a}}){C{\^o}t{\'e}}, {Lugaro}, {Reifarth}, {Pignatari}, {Vil{\'a}gos}, {Yag{\"u}e}, \& {Gibson}}]{cote+19_giugno}
{C{\^o}t{\'e}}, B., {Lugaro}, M., {Reifarth}, R., {et~al.} 2019{\natexlab{a}}, \apj, 878, 156

\bibitem[{{C{\^o}t{\'e}} {et~al.}(2019{\natexlab{b}}){C{\^o}t{\'e}}, {Yag{\"u}e}, {Vil{\'a}gos}, \& {Lugaro}}]{cote+19_dicembre}
{C{\^o}t{\'e}}, B., {Yag{\"u}e}, A., {Vil{\'a}gos}, B., \& {Lugaro}, M. 2019{\natexlab{b}}, \apj, 887, 213

\bibitem[{{Cox} \& {G{\'o}mez}(2002)}]{cox2002}
{Cox}, D.~P. \& {G{\'o}mez}, G.~C. 2002, \apjs, 142, 261

\bibitem[{{D'Antona} \& {Matteucci}(1991)}]{DAntona&Matteucci91}
{D'Antona}, F. \& {Matteucci}, F. 1991, \aap, 248, 62

\bibitem[{{Della Valle} \& {Izzo}(2020)}]{dV&I2020}
{Della Valle}, M. \& {Izzo}, L. 2020, \aapr, 28, 3

\bibitem[{{Desch} {et~al.}(2023){Desch}, {Dunlap}, {Dunham}, {Williams}, \& {Mane}}]{Desch+2023}
{Desch}, S.~J., {Dunlap}, D.~R., {Dunham}, E.~T., {Williams}, C.~D., \& {Mane}, P. 2023, \icarus, 402, 115607

\bibitem[{{Diehl}(2013)}]{Diehl+13}
{Diehl}, R. 2013, Reports on Progress in Physics, 76, 026301

\bibitem[{{Diehl}(2016)}]{Diehl+16}
{Diehl}, R. 2016, in Journal of Physics Conference Series, Vol. 665, Journal of Physics Conference Series (IOP), 012011

\bibitem[{{Diehl} {et~al.}(1995){Diehl}, {Dupraz}, {Bennett}, {Bloemen}, {Hermsen}, {Knoedlseder}, {Lichti}, {Morris}, {Ryan}, {Schoenfelder}, {Steinle}, {Strong}, {Swanenburg}, {Varendorff}, \& {Winkler}}]{Diehl+95}
{Diehl}, R., {Dupraz}, C., {Bennett}, K., {et~al.} 1995, \aap, 298, 445

\bibitem[{{Diehl} \& {Prantzos}(2023)}]{Diehl&Prantzos23}
{Diehl}, R. \& {Prantzos}, N. 2023, arXiv e-prints, arXiv:2303.01825

\bibitem[{{Flynn} {et~al.}(2006){Flynn}, {Holmberg}, {Portinari}, {Fuchs}, \& {Jahrei{\ss}}}]{Flynn+06}
{Flynn}, C., {Holmberg}, J., {Portinari}, L., {Fuchs}, B., \& {Jahrei{\ss}}, H. 2006, \mnras, 372, 1149

\bibitem[{{Ford}(1978)}]{ford78}
{Ford}, H.~C. 1978, \apj, 219, 595

\bibitem[{{GRAVITY Collaboration} {et~al.}(2021){GRAVITY Collaboration}, {Abuter}, {Amorim}, {Baub{\"o}ck}, {Baganoff}, {Berger}, {Boyce}, {Bonnet}, {Brandner}, {Cl{\'e}net}, {Davies}, {de Zeeuw}, {Dexter}, {Dallilar}, {Drescher}, {Eckart}, {Eisenhauer}, {Fazio}, {F{\"o}rster Schreiber}, {Foster}, {Gammie}, {Garcia}, {Gao}, {Gendron}, {Genzel}, {Ghisellini}, {Gillessen}, {Gurwell}, {Habibi}, {Haggard}, {Hailey}, {Harrison}, {Haubois}, {Hei{\ss}el}, {Henning}, {Hippler}, {Hora}, {Horrobin}, {Jim{\'e}nez-Rosales}, {Jochum}, {Jocou}, {Kaufer}, {Kervella}, {Lacour}, {Lapeyr{\`e}re}, {Le Bouquin}, {L{\'e}na}, {Lowrance}, {Lutz}, {Markoff}, {Mori}, {Morris}, {Neilsen}, {Nowak}, {Ott}, {Paumard}, {Perraut}, {Perrin}, {Ponti}, {Pfuhl}, {Rabien}, {Rodr{\'\i}guez-Coira}, {Shangguan}, {Shimizu}, {Scheithauer}, {Smith}, {Stadler}, {Stern}, {Straub}, {Straubmeier}, {Sturm}, {Tacconi}, {Vincent}, {von Fellenberg}, {Waisberg}, {Widmann}, {Wieprecht}, {Wiezorrek}, {Willner}, {Witzel}, {Woillez}, {Yazici}, {Young}, {Zhang},
  \& {Zins}}]{gravity2021}
{GRAVITY Collaboration}, {Abuter}, R., {Amorim}, A., {et~al.} 2021, \aap, 654, A22

\bibitem[{{Grieco} {et~al.}(2012){Grieco}, {Matteucci}, {Pipino}, \& {Cescutti}}]{Grieco+12}
{Grieco}, V., {Matteucci}, F., {Pipino}, A., \& {Cescutti}, G. 2012, \aap, 548, A60

\bibitem[{{Iwamoto} {et~al.}(1999){Iwamoto}, {Brachwitz}, {Nomoto}, {Kishimoto}, {Umeda}, {Hix}, \& {Thielemann}}]{Iwamoto+99}
{Iwamoto}, K., {Brachwitz}, F., {Nomoto}, K., {et~al.} 1999, \apjs, 125, 439

\bibitem[{{Jos{\'e}} \& {Hernanz}(2007)}]{J&H07}
{Jos{\'e}}, J. \& {Hernanz}, M. 2007, Journal of Physics G Nuclear Physics, 34, R431

\bibitem[{{Karakas}(2010)}]{karakas2010}
{Karakas}, A.~I. 2010, \mnras, 403, 1413

\bibitem[{{Kennicutt}(1998)}]{Kennicutt98}
{Kennicutt}, Robert~C., J. 1998, \apj, 498, 541

\bibitem[{{Kobayashi} {et~al.}(2006){Kobayashi}, {Umeda}, {Nomoto}, {Tominaga}, \& {Ohkubo}}]{Kobayashi06}
{Kobayashi}, C., {Umeda}, H., {Nomoto}, K., {Tominaga}, N., \& {Ohkubo}, T. 2006, \apj, 653, 1145

\bibitem[{{Kroupa} {et~al.}(1993){Kroupa}, {Tout}, \& {Gilmore}}]{kroupa1993}
{Kroupa}, P., {Tout}, C.~A., \& {Gilmore}, G. 1993, \mnras, 262, 545

\bibitem[{{Kubryk} {et~al.}(2015{\natexlab{a}}){Kubryk}, {Prantzos}, \& {Athanassoula}}]{KPA15}
{Kubryk}, M., {Prantzos}, N., \& {Athanassoula}, E. 2015{\natexlab{a}}, \aap, 580, A126

\bibitem[{{Kubryk} {et~al.}(2015{\natexlab{b}}){Kubryk}, {Prantzos}, \& {Athanassoula}}]{kubryk2015}
{Kubryk}, M., {Prantzos}, N., \& {Athanassoula}, E. 2015{\natexlab{b}}, \aap, 580, A126

\bibitem[{{Lehner} \& {Howk}(2011)}]{L&H11}
{Lehner}, N. \& {Howk}, J.~C. 2011, Science, 334, 955

\bibitem[{{Limongi} \& {Chieffi}(2006)}]{Limongi+06}
{Limongi}, M. \& {Chieffi}, A. 2006, \apj, 647, 483

\bibitem[{{Marasco} {et~al.}(2012){Marasco}, {Fraternali}, \& {Binney}}]{Marasco+12}
{Marasco}, A., {Fraternali}, F., \& {Binney}, J.~J. 2012, \mnras, 419, 1107

\bibitem[{{Matteucci} \& {Brocato}(1990)}]{M&B90}
{Matteucci}, F. \& {Brocato}, E. 1990, \apj, 365, 539

\bibitem[{{Matteucci} \& {Francois}(1989)}]{matteucci1989}
{Matteucci}, F. \& {Francois}, P. 1989, \mnras, 239, 885

\bibitem[{{Matteucci} \& {Greggio}(1986)}]{matteucci1986}
{Matteucci}, F. \& {Greggio}, L. 1986, \aap, 154, 279

\bibitem[{{Matteucci} {et~al.}(2019){Matteucci}, {Grisoni}, {Spitoni}, {Zulianello}, {Rojas-Arriagada}, {Schultheis}, \& {Ryde}}]{Matteucci+2019}
{Matteucci}, F., {Grisoni}, V., {Spitoni}, E., {et~al.} 2019, \mnras, 487, 5363

\bibitem[{{Matteucci} \& {Recchi}(2001)}]{Matteucci_Recchi2001}
{Matteucci}, F. \& {Recchi}, S. 2001, \apj, 558, 351

\bibitem[{{Meidt} {et~al.}(2009){Meidt}, {Rand}, \& {Merrifield}}]{meidt2009}
{Meidt}, S.~E., {Rand}, R.~J., \& {Merrifield}, M.~R. 2009, \apj, 702, 277

\bibitem[{{Minchev} {et~al.}(2012){Minchev}, {Famaey}, {Quillen}, {Di Matteo}, {Combes}, {Vlaji{\'c}}, {Erwin}, \& {Bland -Hawthorn}}]{minchev2012}
{Minchev}, I., {Famaey}, B., {Quillen}, A.~C., {et~al.} 2012, \aap, 548, A126

\bibitem[{{Nofar} {et~al.}(1991){Nofar}, {Shaviv}, \& {Starrfield}}]{Nofar+91}
{Nofar}, I., {Shaviv}, G., \& {Starrfield}, S. 1991, \apj, 369, 440

\bibitem[{{Nomoto} {et~al.}(1984){Nomoto}, {Thielemann}, \& {Yokoi}}]{NT&Y84}
{Nomoto}, K., {Thielemann}, F.~K., \& {Yokoi}, K. 1984, \apj, 286, 644

\bibitem[{{Pleintinger} {et~al.}(2023){Pleintinger}, {Diehl}, {Siegert}, {Greiner}, \& {Krause}}]{Pleintinger+23}
{Pleintinger}, M. M.~M., {Diehl}, R., {Siegert}, T., {Greiner}, J., \& {Krause}, M. G.~H. 2023, \aap, 672, A53

\bibitem[{{Pl{\"u}schke} {et~al.}(2001){Pl{\"u}schke}, {Diehl}, {Sch{\"o}nfelder}, {Bloemen}, {Hermsen}, {Bennett}, {Winkler}, {McConnell}, {Ryan}, {Oberlack}, \& {Kn{\"o}dlseder}}]{Pluschke+01}
{Pl{\"u}schke}, S., {Diehl}, R., {Sch{\"o}nfelder}, V., {et~al.} 2001, in ESA Special Publication, Vol. 459, Exploring the Gamma-Ray Universe, ed. A.~{Gimenez}, V.~{Reglero}, \& C.~{Winkler}, 55--58

\bibitem[{{Prantzos} {et~al.}(2011){Prantzos}, {Boehm}, {Bykov}, {Diehl}, {Ferri{\`e}re}, {Guessoum}, {Jean}, {Knoedlseder}, {Marcowith}, {Moskalenko}, {Strong}, \& {Weidenspointner}}]{Prantzos+11}
{Prantzos}, N., {Boehm}, C., {Bykov}, A.~M., {et~al.} 2011, Reviews of Modern Physics, 83, 1001

\bibitem[{{Prantzos} \& {Diehl}(1996)}]{Prantzos&Diehl96}
{Prantzos}, N. \& {Diehl}, R. 1996, \physrep, 267, 1

\bibitem[{{Quillen} {et~al.}(2011){Quillen}, {Dougherty}, {Bagley}, {Minchev}, \& {Comparetta}}]{quillen2011}
{Quillen}, A.~C., {Dougherty}, J., {Bagley}, M.~B., {Minchev}, I., \& {Comparetta}, J. 2011, \mnras, 417, 762

\bibitem[{{Quillen} \& {Minchev}(2005)}]{quillen2005}
{Quillen}, A.~C. \& {Minchev}, I. 2005, \aj, 130, 576

\bibitem[{{Roca-F{\`a}brega} {et~al.}(2014){Roca-F{\`a}brega}, {Antoja}, {Figueras}, {Valenzuela}, {Romero-G{\'o}mez}, \& {Pichardo}}]{roca2014}
{Roca-F{\`a}brega}, S., {Antoja}, T., {Figueras}, F., {et~al.} 2014, \mnras, 440, 1950

\bibitem[{{Romano} \& {Matteucci}(2003)}]{Romano&Matteucci03}
{Romano}, D. \& {Matteucci}, F. 2003, \mnras, 342, 185

\bibitem[{{Salpeter}(1955)}]{Salpeter55}
{Salpeter}, E.~E. 1955, \apj, 121, 161

\bibitem[{{Sch{\"o}nfelder} {et~al.}(1984){Sch{\"o}nfelder}, {Diehl}, {Lichti}, {Steinle}, {Swanenburg}, {Deerenberg}, {Aarts}, {Lockwood}, {Webber}, {Macri}, {Ryan}, {Simpson}, {Taylor}, {Bennett}, \& {Snelling}}]{Schonfelder+84}
{Sch{\"o}nfelder}, V., {Diehl}, R., {Lichti}, G.~G., {et~al.} 1984, IEEE Transactions on Nuclear Science, 1, 766

\bibitem[{{Schulreich} {et~al.}(2023){Schulreich}, {Feige}, \& {Breitschwerdt}}]{Schulreich+23}
{Schulreich}, M.~M., {Feige}, J., \& {Breitschwerdt}, D. 2023, \aap, 680, A39

\bibitem[{{Sellwood} \& {Carlberg}(2014)}]{sellwood2014}
{Sellwood}, J.~A. \& {Carlberg}, R.~G. 2014, \apj, 785, 137

\bibitem[{{Siegert} {et~al.}(2023){Siegert}, {Pleintinger}, {Diehl}, {Krause}, {Greiner}, \& {Weinberger}}]{Siegert+23}
{Siegert}, T., {Pleintinger}, M. M.~M., {Diehl}, R., {et~al.} 2023, \aap, 672, A54

\bibitem[{{Spitoni} {et~al.}(2019){Spitoni}, {Cescutti}, {Minchev}, {Matteucci}, {Silva Aguirre}, {Martig}, {Bono}, \& {Chiappini}}]{Spitoni+19}
{Spitoni}, E., {Cescutti}, G., {Minchev}, I., {et~al.} 2019, \aap, 628, A38

\bibitem[{{Spitoni} {et~al.}(2023){Spitoni}, {Cescutti}, {Recio-Blanco}, {Minchev}, {Poggio}, {Palicio}, {Matteucci}, {Peirani}, {Barbillon}, \& {Vasini}}]{Spitoni+23}
{Spitoni}, E., {Cescutti}, G., {Recio-Blanco}, A., {et~al.} 2023, \aap, 680, A85

\bibitem[{{Timmes} {et~al.}(1995){Timmes}, {Woosley}, {Hartmann}, {Hoffman}, {Weaver}, \& {Matteucci}}]{Timmes+95}
{Timmes}, F.~X., {Woosley}, S.~E., {Hartmann}, D.~H., {et~al.} 1995, \apj, 449, 204

\bibitem[{{Vasini} {et~al.}(2022){Vasini}, {Matteucci}, \& {Spitoni}}]{Vasini+22}
{Vasini}, A., {Matteucci}, F., \& {Spitoni}, E. 2022, \mnras, 517, 4256

\bibitem[{{Vasini} {et~al.}(2023){Vasini}, {Matteucci}, {Spitoni}, \& {Siegert}}]{vasini2023}
{Vasini}, A., {Matteucci}, F., {Spitoni}, E., \& {Siegert}, T. 2023, \mnras, 523, 1153

\bibitem[{{Vincenzo} \& {Kobayashi}(2020)}]{vincenzo2020}
{Vincenzo}, F. \& {Kobayashi}, C. 2020, \mnras, 496, 80

\bibitem[{{Wehmeyer} {et~al.}(2023){Wehmeyer}, {L{\'o}pez}, {C{\^o}t{\'e}}, {K. Pet{\H{o}}}, {Kobayashi}, \& {Lugaro}}]{Wehmeyer+23}
{Wehmeyer}, B., {L{\'o}pez}, A.~Y., {C{\^o}t{\'e}}, B., {et~al.} 2023, \apj, 944, 121

\bibitem[{{Winkler}(1994)}]{Winkler+94}
{Winkler}, C. 1994, \apjs, 92, 327

\bibitem[{{Woosley} \& {Weaver}(1995)}]{WW95}
{Woosley}, S.~E. \& {Weaver}, T.~A. 1995, \apjs, 101, 181

\end{thebibliography}

\end{document}